\documentclass[sigconf,nonacm]{acmart}

\begin{document}
\title{Key Management Based on Ownership of Multiple Authenticators in Public Key Authentication}

\author{Koudai Hatakeyama}
\email{hatakeyama@net.ist.i.kyoto-u.ac.jp}
\affiliation{
  \institution{Kyoto University}
  \streetaddress{Yoshida-Honmachi}
  \city{Sakyo}
  \state{Kyoto}
  \country{Japan}
  \postcode{606-8501}
}

\author{Daisuke Kotani}
\email{kotani@media.kyoto-u.ac.jp}
\orcid{0000-0003-4305-8379}
\affiliation{%
  \institution{Kyoto University}
  \streetaddress{Yoshida-Honmachi}
  \city{Sakyo}
  \state{Kyoto}
  \country{Japan}
  \postcode{606-8501}
}

\author{Yasuo Okabe}
\email{okabe@i.kyoto-u.ac.jp}
\orcid{0000-0003-0825-2256}
\affiliation{%
  \institution{Kyoto University}
  \streetaddress{Yoshida-Honmachi}
  \city{Sakyo}
  \state{Kyoto}
  \country{Japan}
  \postcode{606-8501}
}

\begin{abstract}
Public key authentication (PKA) has been deployed in various services to provide stronger authentication to users. 
In PKA, a user manages private keys on her devices called authenticators, and services bind the corresponding public keys to her account. 
To protect private keys, a user uses authenticators which never export private keys outside.
On the other hand, a user regularly uses multiple authenticators like PCs and smartphones.
She replaces some of her authenticators according to their lifecycle, such as purchasing new devices and losing devices. 
It is a burden for a user to register, update and revoke public keys in many services every time she registers new accounts with services and replaces some of her authenticators.
To ease the burden, we propose a mechanism where users and services manage public keys based on the owner of authenticators and users can access services with PKA using any of their authenticators.
We introduce a key pair called an Ownership Verification Key (OVK), which consists of the private key (OVSK) and the corresponding public key (OVPK). 
All authenticators owned by a user derive the same OVSK from the pre-shared secret called the seed.
Services verify the ownership of the authenticators using the corresponding OVPK to determine whether binding the requested public key to her account. 
To protect user privacy while maintaining convenience, authenticators generate a different OVK for each service from the seed independently. 
We demonstrate the feasibility through the Proof of Concept implementation, show that our proposed mechanism achieves some security goals, and discuss how the mechanism mitigates threats not completely handled.
\end{abstract}


\maketitle

\section{Introduction} \label{sec:hajimeni}
Password authentication has long been essential for remote access to services that require user authentication \cite{password-security-a-case-history}, but there remain many problems with it \cite{quest-to-replace-passwords}. 
Phishing, for example, is a problem caused by asking users to present their passwords every time they use a service \cite{anti-phishing}.
It is also a problem that users use the same password for many services or use short passwords because they don't want to remember many complex passwords \cite{users-are-not-the-enemy, password-management-strategies-for-online-accounts}.

Public key authentication is an alternative to password authentication for stronger authentication.
Public key authentication assumes that only a user has a private key, and a service has the corresponding public key.
A service authenticates a user in the following three steps.
First, the service sends a random number called a challenge to the user.
Second, she signs the challenge by the private key corresponding to the public key registered with the service.
Third, the service verifies the signature by any one of the public keys bound to her account.

Services have to manage the binding of public keys to a user's account.
Services trust this binding based on several models, where trusted third parties ensure the binding (e.g., Individual Number Card in Japan \cite{my-number} and WebPKI \cite{rfc5280}) and where they receive keys from users directly via trusted channels (e.g., FIDO \cite{how-fido-works} and SSH \cite{rfc4252}).
This study will focus on the last model.

Users have to manage private keys corresponding to registered public keys on their devices.
We call these devices authenticators.
Authenticators, such as Yubikey \cite{yubico-product-doc} and Keychain \cite{apple-keychain}, store key pairs in secure storage where corresponding private keys cannot be exported nor easily accessed by the outside of the authenticator.
Operations using keys stored in secure storage require local authentication by authenticators, like PIN or biometrics.
This study assumes that users use authenticators having secure storage.

An authenticator has a mechanism called attestation \cite{fido-attestation} that proves that an operation is done surely by the authenticator.
An attestation includes information about the manufacturer of the authenticator which generates the attestation, the model name of the authenticator, and results of the operation done by the authenticator, such as the public key of a generated key pair.
Attestations are signed by the attestation key embedded in an authenticator by its manufacturer so that services can validate whether received attestations are generated by the authenticator.
During account registration, an authenticator sends a public key, an attestation about the public key, and the certificate of the attestation key.
Services can determine the trustworthiness of the received public key and the authenticator that stores the corresponding private key by verifying the attestation with the certificates received from the manufacturer of the authenticator.

Public key authentication (PKA) is stronger than password authentication for the following reason.
First, PKA is resistant to data breaches on services because attackers cannot sign in to services with only public keys.
Second, PKA can be phishing resistant when an authenticator verifies whether the requested service is the same as the service accessed previously without interaction with a user.
Lastly, users don't have to use weak private keys for convenience, because authenticators, not users, remember and manage these keys.
Besides, Malicious services cannot correlate their account using registered public keys because authenticators generate different public keys for each service.

However, public key authentication has the problem that users can only use authenticators storing private keys corresponding to registered public keys when accessing services.
Given the following two concerns, it is a burden for users to register, update and revoke public keys in many services.
First, users usually have multiple authenticators such as smartphones, PCs, and tablets.
They have to register multiple public keys with their authenticators, but simply registering public keys with each of all authenticators annoys users \cite{fido-usability}.
Second, users add, replace, and throw away their authenticators according to the lifecycles of the authenticators.
Once such an event occurs, users need to update and revoke registered public keys in many services.
Currently, users can manage registered public keys on services via an authenticated session.
If an attacker steals an authenticator and revokes the public keys of authenticators held by a legitimate user before the user revokes the public key of the stolen authenticator, the user may become inaccessible.

The purpose of this study is that users can access services with public key authentication using any owned authenticators without explicitly registering public keys.
To realize this purpose, we propose the mechanism where users and services manage public keys based on the owner of authenticators storing the corresponding private keys.
We introduce a key pair, called an Ownership Verification Key (OVK).
A user proves the ownership of authenticators by the private key of an OVK (Ownership Verification Secret Key; OVSK).
A service verifies the possession of the authenticators by the public key of the OVK (Ownership Verification Public Key; OVPK).
All authenticators owned by a user can derive an OVSK from a seed pre-shared among them.
A service manages the corresponding OVPK by binding it to her account.
A service binds public keys signed by the OVSK to her account if verification by the OVPK is successful.
To protect user privacy while maintaining convenience, authenticators generate a different OVK for each service from the seed independently. 
Users and services update OVKs according to the lifecycles of users' authenticators.
When a user changes a set of her authenticators, she updates an OVSK, and services update an OVPK bound to her accounts.

The main contribution of this paper is that users and services can manage public keys based on the owner of the authenticators storing the corresponding private keys to facilitate their key management in public key authentication.
We implemented the Proof of Concept and confirmed that key management works as expected for typical use cases.
We analyzed the proposed mechanism to find threats with threat modeling and evaluated what measures our proposal takes against the found threats.
We confirmed that our proposal achieves some security goals such as that services cannot correlate accounts and can correctly bind public keys to accounts.
We discussed how our proposal mitigates threats for which measures are not sufficient. 

The following is the structure of this paper.
Section \ref{sec:haikei} describes related work.
Section \ref{sec:teian} describes the key management using OVK.
Section \ref{sec:poc} describes the implementation of the Proof of Concept and use cases using the Proof of Concept.
Section \ref{sec:eval} describes evaluation with threat modeling.
Section \ref{sec:kousatsu} discusses the proposal.
Finally, Section \ref{sec:conclusion} summarizes this paper.

\section{Related Work} \label{sec:haikei}
Nishimura \cite{ntt-owner-identity} proposes sharing private keys among authenticators that users own.
Authenticators verify the owner of other authenticators to determine whether sharing private keys or not.
To verify the owner, a trusted third party issues certificates to authenticators.
However, this approach weakens the authentication level of public key authentication because authenticators export private keys from secure storage.
This approach also weakens the trustworthiness of a registered public key because services cannot verify attestations.
Besides, reliance on a trusted third party has a large management cost and the impact like a certificate authority in WebPKI if it becomes untrustworthy.

James \cite{conners-19-lets-authenticate} introduces certificate chains like TLS client authentication to FIDO public keys so that FIDO is capable of registering multiple authenticators and recovering accounts.
When a user registers a new public key generated in her authenticator with a service, she requests a certificate authority to issue the certificate binding the public key to her account of the service.
The authenticator sends the certificate to the service and the service can verify the owner of the public key by checking the subject of the certificate.
However, this approach has the problem that it is not clear how a certificate authority authenticates users using multiple authenticators in addition to the same problem due to a trusted third party as \cite{ntt-owner-identity}.

Oogami \cite{fido-multi-registration} proposes the mechanism in which users register a new FIDO public key of an authenticator via authenticated sessions established by the registered public key of other authenticators.
Public keys have high assurance because users use registered authenticators every time users register a new public key of an authenticator. 
However, users have to keep multiple authenticators at the same time when registering a new public key, so that users cannot register a new public key when they have only unregistered authenticators.

Frymann \cite{yubico-webauthn-account-recovery} and Lundberg \cite{yubico-webatuhn-account-recovery-impl} propose a mechanism for account recovery when losing registered authenticators that users use daily.
In this mechanism, a user has two authenticators.
One is for daily use by users, called the main authenticator.
The other is for backup use by users, called the backup authenticator.
The user deposits the backup authenticator in a vault.
The main authenticator receives the seed for deriving public keys from the backup authenticator in advance.
On behalf of the backup authenticator, the main authenticator generates a different public key of the backup authenticator for each service and registers the public key whose corresponding private key the backup authenticator can only derive.
The user can access services with the backup authenticator when losing the main authenticator.
As a result, this approach prevents services from correlating their account based on the registered public keys.
However, services cannot verify the attestation of the public key of the backup authenticator during registration.
Besides, when attackers gain control of the backup authenticator, they sign in with the backup authenticator and can revoke the public key of the main authenticator, and the user cannot sign in with the main authenticator.

Identity Federations using OpenID Connect \cite{oidc-core} or SAML \cite{saml-v2} allow users to reduce the number of services where users register public keys.
However, users still register public keys with several services.
Besides, there are also privacy issues where the service authenticating users, called an Identity Provider, can know what services they are using.

\section{Key Management with an Ownership Verification Key} \label{sec:teian}
\subsection{Overview}
We propose the mechanism where a user and a service manage keys for authentication based on a public key cryptographic key pair called an Ownership Verification Key (OVK).
An OVK is derived by all authenticators of a user to prove that the private key corresponding to the public key to be registered is stored in her owned authenticator.
The public key of the OVK (Ownership Verification Public Key; OVPK) is registered with the service via the trusted channel established when registering a new account.
The service binds the OVPK to her account.
The private key of the OVK (Ownership Verification Secret Key; OVSK) is used for signing the public key to be registered.
The service binds the public key to her account if verification by the OVPK is successful.

\begin{figure}[htb]
  \centering
  \includegraphics[width=\columnwidth]{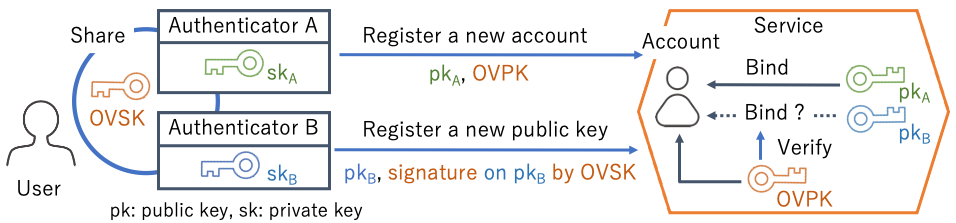}
  \caption{Registering Public Keys Using An OVK}
  \label{fig:ovk-gaiyou-1}
\end{figure}

Fig.\ \ref{fig:ovk-gaiyou-1} illustrates how a user registers a public key when she has two authenticators (A and B).
She shares an OVSK among Authenticators A and B in advance.
When she registers a new account using Authenticator A, she attaches a public key for Authenticator A and an OVPK to the service.
Then she seamlessly registers a new public key for Authenticator B by signing the public key with the OVSK whose corresponding OVPK has been already registered with Authenticator A.
The service verifies the signature by the registered OVPK and, if succeeded, binds the public key to her account.

Users and services update OVKs according to the lifecycles of users' authenticators.
When a user changes a set of her authenticators, she updates an OVK in her all authenticators and notifies services of updating the OVPK.
To make an updating message, registered authenticators sign the new OVPK to be updated by the previous OVSK whose corresponding OVPK is now registered with services.
Services update an OVPK bound to her account based on the most trustworthy updating message and re-bind public keys to her account based on the new OVPK.
A user can still sign in with authenticators that have notified services of the new OVPK.
To invalidate the authenticators that are no longer in use, services revoke the public key corresponding to the authenticators that are not bounded to the new OVPK.

\begin{figure}[htb]
  \centering
  \includegraphics[width=\columnwidth]{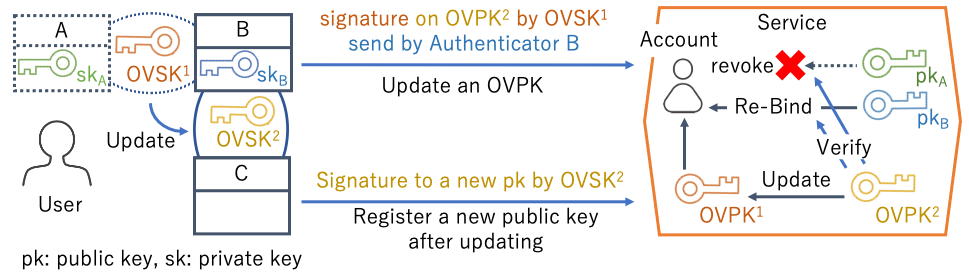}
  \caption{Updating an OVK}
  \label{fig:ovk-gaiyou-2}
\end{figure}

Fig,\ \ref{fig:ovk-gaiyou-2} illustrates how a user updates an OVK when she has had two authenticators (A and B) and registered them with a service and replaces Authenticator A with Authenticator C because of losing Authenticator A.
After losing Authenticator A, she shares a new OVSK between Authenticators B and C.
Authenticator B derives the new OVK ($OVK^2$) and signs the new OVPK ($OVPK^2$) by the previous OVSK ($OVSK^1$) to make an updating message.
Authenticator B notifies the service of the new OVPK ($OVPK^2$) by sending the updating message.
The service evaluates the message received from Authenticator B as the most trustworthy and binds the new OVPK to her account.
It re-binds the public key for Authenticator B to her account because B has sent the most trustworthy message and revokes the public key for Authenticator A because A has sent no messages.
Then she can seamlessly register a new public key for Authenticator C by signing the public key with the new OVSK ($OVSK^2$) whose corresponding OVPK ($OVPK^2$) has been already registered by Authenticator B.

The following is the structure of this section.
Section \ref{sec:teian:ovk-derive} describes how to derive an OVK among authenticators.
We introduce a pre-shared secret called a seed to derive an OVK.
Section \ref{sec:teian:seed-sharing} describes how to share a seed among authenticators.
Section \ref{sec:teian:ovk-trust} describes how services verify the trustworthiness of an OVK requested to be registered.
Section \ref{sec:teian:ovk-update} describes how to update an OVK after sharing a new seed.

\subsection{Deriving an OVK from a Shared Secret} \label{sec:teian:ovk-derive}
In this section, we describe how to derive an OVK from the pre-shared secret, called the seed.
We assume that the seed has been shared among all authenticators owned by the same user.
We also explain how to register public keys using an OVK.

\subsubsection{Requirement} \label{sec:teian:ovk-derive:youken}
We define the requirements in such a way that our proposal does not interfere with what public key authentication described in Section \ref{sec:hajimeni}, which we call PKA, can achieve during public key registration \cite{fido-sec-ref, fido-privacy-principles}.

First, our proposed method should not rely on trusted third parties for proving the owner of authenticators except for verifying attestations and establishing secure channels.
In PKA, users can register public keys via a trusted channel established when registering a new account or established by registered authenticators.

Second, our proposed method must prevent services from correlating their account by using the proof of the owner of authenticators.
In PKA, users can register different public keys with each service to protect user privacy against services seeking to correlate their accounts based on registered public keys.

Third, in our proposed method, services should verify the attestation of the public key requested to be registered.
Services can calculate the trustworthiness of the public key by verifying the attestation generated by the authenticator that has the corresponding private key.

Finally, our proposed method should minimize the number of times a user operates multiple authenticators at the same time for convenience.
Operating multiple authenticators at the same time whenever registering a new public key annoys a user.

\subsubsection{Deriving an OVK} \label{sec:teian:ovk-derive:process}
\begin{figure}[htb]
  \centering
  \includegraphics[width=\columnwidth]{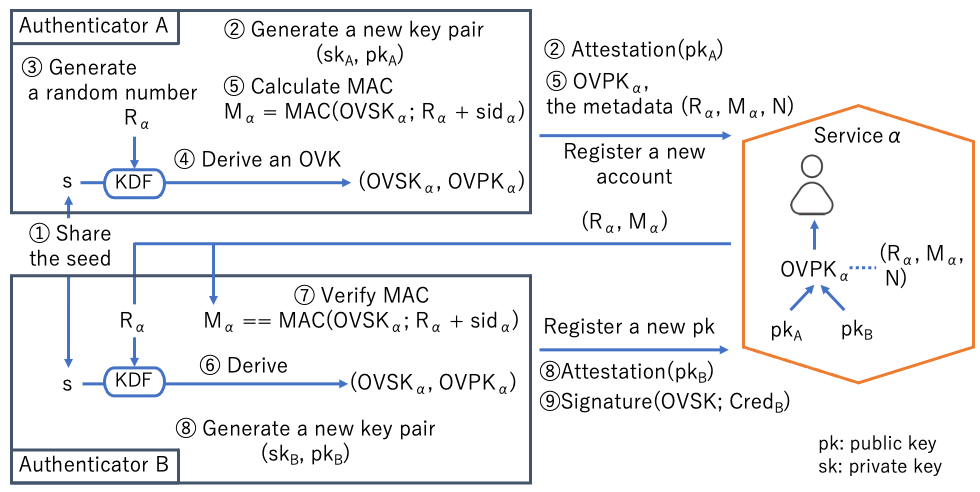}
  \caption{Deriving an OVK from the shared seed}
  \label{fig:ovk-derive}
\end{figure}

We explain our proposal using Fig.\ \ref{fig:ovk-derive}, which shows a case where a user has two authenticators, A and B.
A user registers a new account with service $\alpha$ using Authenticator A at first, and then she accesses the service using Authenticator B.
Note that we assume that messages between authenticators and the service have protected in terms of service authentication, confidentiality, and integrity (e.g., via TLS).

The two authenticators agree in advance on the following parameters and the identifier of service $\alpha$ ($sid_\alpha$).

\begin{itemize}
  \item \texttt{s}: The seed value shared among authenticators (\textcircled{\scriptsize 1})
  \item \texttt{N}: The number of authenticators sharing the same seed (equal to the number of her authenticators) 
  \item The public key cryptographic algorithm for an OVK
  \item \texttt{KDF}: The key derivation function that takes a seed and a random value as inputs and outputs pseudorandom numbers of the length required for an OVSK
  \item \texttt{MAC}: The message authentication code function that takes an OVSK as a key
\end{itemize}

First, the user registers a new account with service $\alpha$ using Authenticator A.
Authenticator A generates a new key pair ($sk_A, pk_A$) and an attestation of the public key (\textcircled{\scriptsize 2} in Fig.\ \ref{fig:ovk-derive}).
At the same time, Authenticator A derives an OVPK and the corresponding metadata and registers them in addition to the public key ($pk_A$) with service $\alpha$.
The derivation consists of the following three steps.

\begin{enumerate}
  \renewcommand{\labelenumi}{\textcircled{\scriptsize \theenumi}}
  \setcounter{enumi}{2}
  \item Generate a random number ($R_\alpha$).
  \item Calculate an OVSK ($OVSK_\alpha = \mathtt{KDF}(\mathtt{s}, R_\alpha)$) and the corresponding OVPK ($OVPK_\alpha$).
  If the authenticator cannot derive validate OVSK using the random number ($R_\alpha$), start over from \textcircled{\scriptsize 3}.
  \item Register $OVPK_\alpha$ and the corresponding metadata consisting the following three values.
    \begin{itemize}
      \item $R_\alpha$: The generated random number
      \item $M_\alpha$: The message authentication code ($\mathtt{MAC}(OVSK_\alpha, R_\alpha + sid_\alpha)$)
      \item \texttt{N}
    \end{itemize}
\end{enumerate}
Now, the user has registered a new account with service $\alpha$.
The service binds the public key ($pk_A$), $OVPK_\alpha$, and the corresponding metadata ($R_\alpha, M_\alpha, N$) to the new account.

Second, the user access service $\alpha$ using unregistered Authenticator B.
When the service returns a challenge for public key authentication by replying to an authentication request, it also returns the metadata ($R_\alpha$ and $M_\alpha$).
Authenticator B starts on seamless registration of a new public key because B has no public key for signing in to service $\alpha$.
Authenticator B generates a new key pair ($sk_B, pk_B$) and the attestation of the public key (\textcircled{\scriptsize 8} in Fig.\ \ref{fig:ovk-derive}).
Authenticator B signs the public key ($pk_B$) by $OVSK_\alpha$ so that service $\alpha$ verifies whether the owner of Authenticator B storing the corresponding private key ($pk_B$) is the same as the owner of registered authenticators.
To derive $OVSK_\alpha$ from the received metadata, the following two steps are performed.

\begin{enumerate}
  \renewcommand{\labelenumi}{\textcircled{\scriptsize \theenumi}}
  \setcounter{enumi}{5}
  \item Derive $OVK_\alpha$ using the received metadata $R_\alpha$ in the same way as \textcircled{\scriptsize 4}.
  \item Verify the received metadata $M_\alpha$ using the derived $OVSK_\alpha$. 
  If this verification failed, the derived OVSK or the received metadata is not for service $\alpha$.
\end{enumerate}

Service $\alpha$ registers the public key ($pk_B$) if the attestation (\textcircled{\scriptsize 8}) and the signature (\textcircled{\scriptsize 9}) is valid and the number of the registered public keys is not more than \texttt{N}.

\subsubsection{Different OVKs per Service} \label{sec:teian:ovk-derive:ovk-per-svc}
\begin{figure}[htb]
  \centering
  \includegraphics[width=\columnwidth]{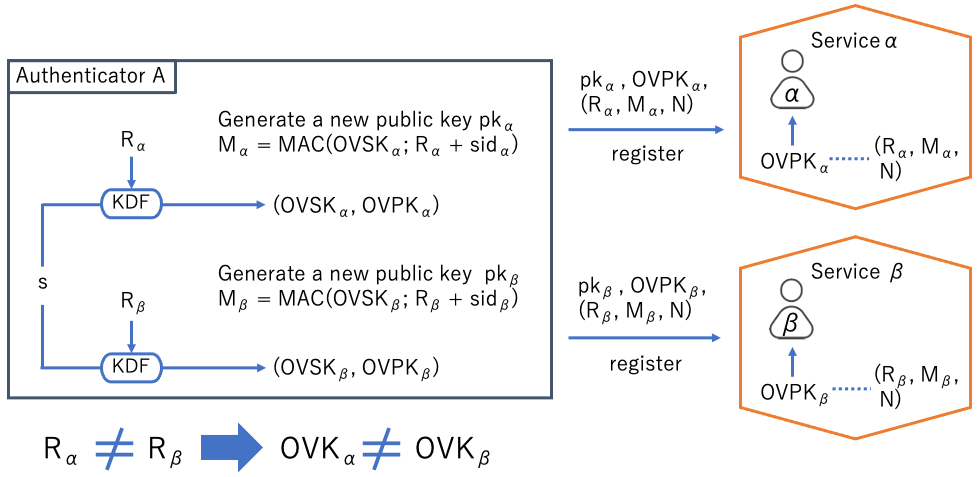}
  \caption{Deriving different OVKs per services}
  \label{fig:ovk-derive-per-svc}
\end{figure}

Authenticators can derive different OVKs per service because of generating different random numbers (\texttt{R}) per service.
Fig.\ \ref{fig:ovk-derive-per-svc} shows a situation where the user registers with Services $\alpha$ and $\beta$.
It is impossible to determine the seed value of an OVSK because of the properties of a key derivation function (\texttt{KDF}).
By using a different random number for each service ( $R_\alpha$ for Service $\alpha$ and $R_\beta$ for Service $\beta$), authenticators can register unlinkable OVPKs with different services.

Authenticators only need to remember the value of the seed. This is because authenticators store random numbers \texttt{R} to services in a verifiable format. 
Moreover, even though the number of registered services increases, a user does not have to operate multiple authenticators to share a new OVSK.
This is convenience for a user.

\subsection{Sharing a Seed among Authenticators} \label{sec:teian:seed-sharing}
In this section, we describe how to share a seed among authenticators of a user.

\subsubsection{Requirement} \label{sec:teian:seed-sharing:youken}
A user operates multiple authenticators and makes them communicate to share a seed.
There are various kinds of short-range communication protocols (e.g., Bluetooth, NFC, and generating and reading QR codes), each of which has its different characteristics in the security of the communication channel.
We define the following requirement to be independent of specific communication protocols.

\begin{itemize}
  \item Assuming no security features of communication channels
\end{itemize}

This requires that attackers cannot calculate a seed using only the information that authenticators send to the channel (resistance to eavesdropping).
This also requires that authenticators can validate whether the received information is generated by the legitimate authenticator to be resistant to tampering.

\subsubsection{Two Authenticators} \label{sec:teian:seed-sharing:2-party}
\begin{figure}[htb]
  \centering
  \includegraphics[width=\columnwidth]{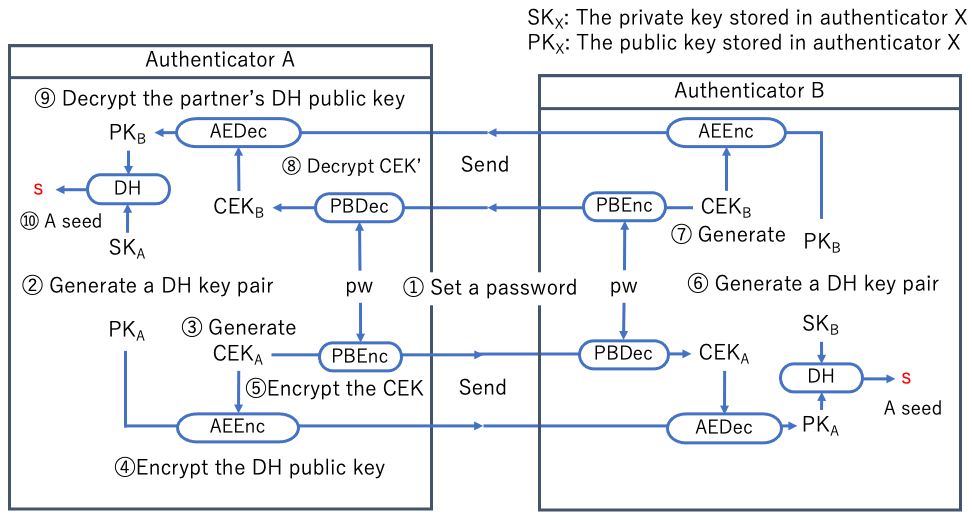}
  \caption{Sharing a seed between two authenticators}
  \label{fig:seed-sharing-2-party}
\end{figure}

Fig.\ \ref{fig:seed-sharing-2-party} shows the case where a user has two authenticators.
Two authenticators agree on the same seed based on the Diffie-Hellman key agreement algorithm.
They encrypt DH public keys using an authenticated encryption based on a password set by the user to ensure the confidentiality of DH public keys and verify the authenticity.

The authenticators agree on the following parameters in advance.

\begin{itemize}
  \item \texttt{pw}: The password set by the user (\textcircled{\scriptsize 1})
  \item \texttt{DH}: A Diffie-hellman key agreement algorithm
  \item The list of authenticated encryption algorithms for encrypting DH public keys (Each algorithm has assigned an identifier)
  \item The list of password-based encryption algortihms (Each algorithm has assigned an identifier)
\end{itemize}

First, the user operates Authenticator A.

\begin{enumerate}
  \renewcommand{\labelenumi}{\textcircled{\scriptsize \theenumi}}
  \setcounter{enumi}{1}
  \item Generate a DH key pair ($SK_A, PK_A$).
  \item Generate a random number called a Content Encryption Key ($CEK_A$).
  \item Encrypt the DH public key ($PK_A$) using $CEK_A$. 
  Authenticator A determines the authenticated encryption algorithm from the list.
  \item Encrypt $CEK_A$ using the password (\texttt{pw}).
  Authenticator A determines the password-based algorithm from the list.
\end{enumerate}
Authenticator A sends the generated ciphertexts and the algorithm identifiers (at \textcircled{\scriptsize 4} and \textcircled{\scriptsize 5}) to Authenticator B.

In the same way, Authenticator B generates a DH key pair ($(SK_B, PK_B)$ at \textcircled{\scriptsize 6}) and a random number ($CEK_B$ at \textcircled{\scriptsize 7}), encrypts the DH public key ($PK_B$) using $CEK_B$, and encrypts $CEK_B$ using the password (\texttt{pw}).
Note that $CEK_B$ is not necessarily the same as $CEK_A$ generated by Authenticator A.

Authenticator A receives the ciphertexts from Authenticator B.
\begin{enumerate}
  \renewcommand{\labelenumi}{\textcircled{\scriptsize \theenumi}}
  \setcounter{enumi}{7}
  \item Decrypt a received CEK ($CEK_B$) using the password (\texttt{pw}).
  Authenticator A identifies the password-based algorithm by the received identifier.
  \item Decrypt a received DH public key ($PK_B$) using the decrypted CEK ($CEK_B$).
  Authenticator A identifies the authenticated encryption algorithm by the received identifier.
  \item Agree the same seed using the Diffie-hellman key agreement algorithm.
\end{enumerate}

\subsubsection{Three or More Authenticators}
\begin{figure}[htb]
  \centering
  \includegraphics[width=\columnwidth]{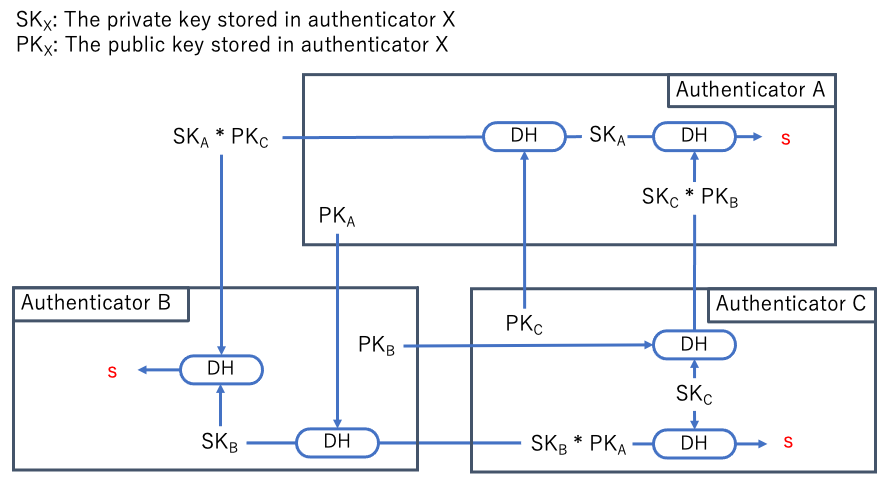}
  \caption{Sharing a seed among three authenticators}
  \label{fig:seed-sharing-3-party}
\end{figure}

When a user has three or more authenticators, the authenticators share a seed like the situation in Fig.\ \ref{fig:seed-sharing-3-party}.
Fig.\ \ref{fig:seed-sharing-3-party} uses the algorithm \cite{multi-party-diffie-hellman}.
All authenticators agree on the following parameters in advance in addition to the agreement for two authenticators.

\begin{itemize}
  \item Each authenticator identifier (These identifiers are temporary identifiers used only to share a seed)
  \item The partner authenticator identifier of each authenticator (Each authenticator receives calculated DH public keys from the same authenticator, called the partner authenticator, every step. The user assigns identifiers without overlap.)
\end{itemize}

The user operates authenticators according to the following steps.
In each step, encryption means that an authenticator generates a CEK, encrypts a DH public key using the CEK, and encrypts the CEK using the password set by a user.

\begin{description}
  \item[Step1] Generate a DH key pair on each authenticator.
  \item[Step2] \begin{itemize}
    \item Authenticator A sends the DH public key ($PK_A$) to Authenticator B with encryption.
    \item B sends the DH public key ($PK_B$) to C with encryption.
    \item C sends the DH public key ($PK_C$) to A with encryption.
  \end{itemize}
  \item[Step3] \begin{itemize}
    \item Authenticator A sends the calculated DH public value ($SK_A * PK_C$) using its DH private key ($SK_A$) and received DH public key ($PK_C$) to Authenticator B with encryption.
    \item B sends the calculated DH public value ($SK_B * PK_A$) using its DH private key ($SK_B$) and received DH public key ($PK_A$) to C with encryption.
    \item C sends the calculated DH public value ($SK_C * PK_B$) using its DH private key ($SK_C$) and received DH public key ($PK_B$) to A with encryption.
  \end{itemize}
  \item[Step4] \begin{itemize}
    \item Authenticator A calculates the DH public value ($SK_A * (SK_C * PK_B)$) and agrees the same seed.
    \item B calculates the DH public value ($SK_B * (SK_A * PK_C)$) and agrees the same seed.
    \item C calculates the DH public value ($SK_C * (SK_B * PK_A)$) and agrees the same seed.
  \end{itemize}
\end{description}

When a user has more than three authenticators ($N$: the number of her authenticators), she processes the above Step1 to Step $N+1$ with repeating the above Step3.

\subsection{Verifying the Trustworthiness of an OVK} \label{sec:teian:ovk-trust}
Because a service binds public keys to an account by an OVK, the trustworthiness of public keys can never be higher than the trustworthiness of the OVK.
We propose how a service verifies the trustworthiness of an OVK.

A service can evaluate the trustworthiness of an OVK using the following two criteria.

\begin{description}
  \item[Criterion1] Whether an OVK is derived as described in Section \ref{sec:teian:ovk-derive}
  \item[Criterion2] Whether a seed is securely stored in all authenticators
\end{description}

The proposed verification mechanism depends on the attestation mechanism that authenticators already have.
Authenticators send an OVPK as well as the attestation of the OVPK at  \textcircled{\scriptsize 5} on Fig.\ \ref{fig:ovk-derive}.

\begin{figure}[htb]
  \centering
  \includegraphics[width=\columnwidth]{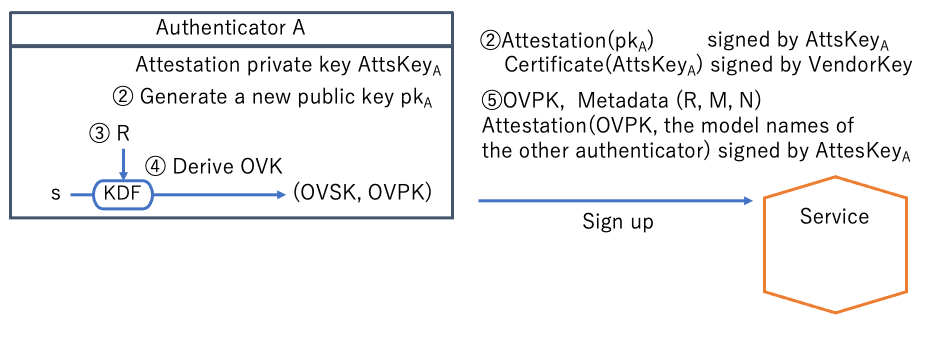}
  \caption{Sending the attestation of an OVPK}
  \label{fig:ovk-attestation}
\end{figure}

Fig.\ \ref{fig:ovk-attestation} extracts the registration flow of an OVPK from Fig.\ \ref{fig:ovk-derive} and describes more details about attestations of public keys.
An attestation private key ($AttsKey_A$) is embedded in an authenticator by its manufacturer. 
A certificate for the attestation public key ($Certificate(AttsKey_A)$) is issued by the manufacturer.
The authenticator signs the public key generated at \textcircled{\scriptsize 2} and the information about the public key by using $AttsKey_A$ and sends them to a service. 

The authenticator also signs a derived OVPK by $AttsKey_A$ to notify the service that the OVPK is derived from the seed stored in the authenticator as described in Section \ref{sec:teian:ovk-derive:process}.
The service can verify the attestation of the OVPK based on the trusted policy about what authenticators comply with Section \ref{sec:teian:ovk-derive}.
The service can validate Criterion1.

\begin{figure}[htb]
  \centering
  \includegraphics[width=\columnwidth]{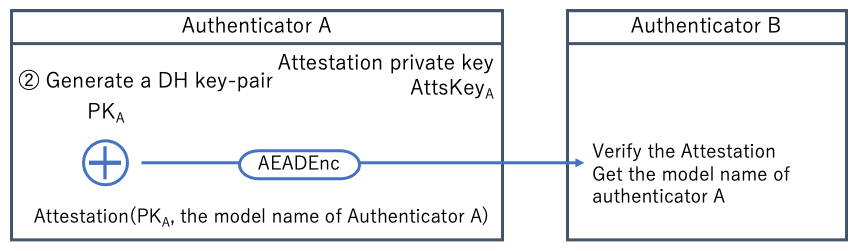}
  \caption{Attestation when sharing a seed}
  \label{fig:seed-sharing-attestation}
\end{figure}

Moreover, an attestation of an OVPK contains the OVPK itself and model names of the other authenticators sharing the same seed.
An authenticator gets the model names of the other authenticators by receiving the attestation of their DH public key while sharing the seed at \textcircled{\scriptsize 2} on Fig.\ \ref{fig:seed-sharing-2-party} (Fig.\ \ref{fig:seed-sharing-attestation}).
A service can verify whether they store the seed securely based on the trusted policy about what authenticator model has secure storage and stores the seed in the storage.
The service can validate Criterion2.

\subsection{Re-sharing a New Seed and Updating an OVK} \label{sec:teian:ovk-update}
A user updates a set of her authenticators according to lifecycles of the authenticators, such as theft or loss.
We propose a mechanism where a user can revoke an OVPK registered with a service and update a new OVPK in the service.

\subsubsection{Assumption} \label{sec:teian:ovk-update:assumption}
We assume the following for this proposal.

\begin{enumerate}
  \item Attackers can operate the seed and the private keys corresponding to registered public keys stored in a stolen authenticator.
  \item It takes time for attackers to gain control of a stolen authenticator. 
\end{enumerate}

Assumption 2 is reasonable when authenticators protect the seed and private keys by local authentication like PIN or biometric.
In Section \ref{sec:kousatsu:update}, we consider the case where authenticators have no local authentication, or where local authentication is immediately passed.

\subsubsection{Overview}
Authenticators share a new seed as described in Section \ref{sec:teian:seed-sharing}.
They hold the previous seed along with a new one without erasing the previous one.
They notify a service of updating an OVK by sending an updating message described in Section \ref{sec:teian:ovk-update:update-message} when a user signs in for the first time after re-sharing the new seed.
The service that receives updating messages waits for some period (OVK migration period) and accepts the new OVPK from the most trustworthy updating message.
A service calculates the trustworthiness of each updating message in the way described in Section \ref{sec:teian:ovk-update:eval-trust}.
The service re-binds public keys to the user's account by verifying with the new OVPK and revokes the public key bound only to the previous OVPK.

\subsubsection{An Updating Message for a New OVK} \label{sec:teian:ovk-update:update-message}
\begin{figure}[htb]
  \centering
  \includegraphics[width=\columnwidth]{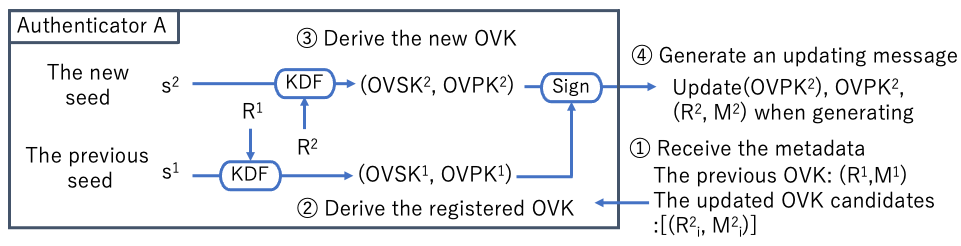}
  \caption{Generating an updating message for a new OVK}
  \label{fig:ovk-update}
\end{figure}

We describe how an authenticator generates an updating message for a new OVK in Fig.\ \ref{fig:ovk-update}.

\begin{enumerate}
  \renewcommand{\labelenumi}{\textcircled{\scriptsize \theenumi}}
  \item The authenticator receives the metadata from the service to derive OVKs from seeds.
  There are two kinds of received metadata. 
  The first one is the metadata ($R^1, M^1$) for the previously registered OVK ($OVK^1$).
  The second one is the metadata ($[(R^2_i, M^2_i)]$) for OVK ($OVK^2$) candidates that other authenticators have registered as new OVKs.
  The authenticator receives the second one as a list because attackers can also generate a malicious updating message derived from a seed by using a stolen authenticator.
  \item The authenticator derives the previously registered OVK ($OVK^1$) from the received metadata ($R^1, M^1$) and the previous seed ($s^1$) as described in Section \ref{sec:teian:ovk-derive}.
  The authenticator uses the same key derivation function (\texttt{KDF}) as the one used in Section \ref{sec:teian:ovk-derive}.
  The authenticator verifies whether the derived OVK ($OVK^1$) is legitimate by comparing calculated MAC value with $M^1$.
  If the verification fails, the authenticator aborts this update process.
  \item The authenticator derives the new OVK ($OVK^2$) from the received metadata ($[(R^2_i, M^2_i)]$) and the newly shared seed ($s^2$).
  If the metadata is not an empty list, the authenticator derives the new OVK from the legitimate metadata ($ (R^2, M^2) = (R^2_l, M^2_l)$) with which the authenticator can verify the MAC value ($M^2_l == MAC(s^2, R^2_l + sid)$).
  If the metadata is an empty list or the metadata has no legitimate metadata, the authenticator generates a new random value ($R^2$) and then derives a new OVK.
  Note that the previously registered OVK ($OVK^1$) is not used to derive the new OVK ($OVK^2$), but used to generate an updating message.
  \item The authenticator signs the new OVPK ($OVPK^2$) by the OVSK ($OVSK^1$) corresponding to the previously registered OVPK ($OVPK^1$).
  The authenticator sends the signature as an updating message when signing in.
  Because the updating message is signed by the private key corresponding to the registered public key, the service can identify the authenticator sending the updating message.
\end{enumerate}

In the above explanation, we assume that authenticators have two shared seeds.
However, authenticators may have more than two seeds because users change their authenticators many times.
Users replace their authenticators when purchasing new devices and may lose their authenticators more than once.
We explain that authenticators can generate the correct updating message even when they have more than two seeds.
Authenticators select the latest seed as the new seed.
As the seed corresponds to registered OVPK, authenticators can select the seed successfully by verifying the MAC value of the received metadata.
From the above, authenticators can send a legitimate updating message to services even when they have more than two seeds.

\subsubsection{Evaluating the Trustworthiness of an Updating Message} \label{sec:teian:ovk-update:eval-trust}
When a service receives an updating message from a registered authenticator, it enters the OVK migration period.
In this migration period, no authenticators can register a new public key by the registered OVK.
If the same updating message comes from more than half of the registered authenticators during the period, the service trusts the message.
Otherwise, the service trusts the updating message sent from the most registered authenticators at the end of the period.
If there is more than one message sent by the most registered authenticators, the service trusts the earliest received message.

\subsubsection{Reducing the Number of Seed Held by Authenticators}
As described in Section \ref{sec:teian:ovk-derive:ovk-per-svc}, an authenticator can derive many different OVKs for different services from a seed.
An authenticator can update OVKs multiple times without consuming a lot of storage space.
However, the secure storage space that an authenticator has is limited.
Therefore, we propose two methods for limiting the number of seeds that an authenticator holds.

\begin{enumerate}
  \item Set a limit on the number of seeds that an authenticator holds.
  If the number of seeds exceeds the limit, it deletes the oldest seed with the consent of the user.
  \item Set an expiration date for a seed.
  If an authenticator has a seed that is about to expire, it prompts a user to share a new seed and update OVKs.
  It deletes any seed that has expired.
\end{enumerate}

The first method is easier to implement because a user decides whether to delete seeds.
The first method does not require a user to renew OVKs periodically, even though she continues to use the same authenticators.
On the other hand, in the second method, since a seed has an expiration date, OVKs also have the same expiration date.
This means that a service can know when to update a seed.
In the second method, a service can send security notifications to reduce operational risks such as forgetting to update OVKs.

Both have their advantages, and both can reduce disadvantages by setting limits to a large value in the first method or a longer expiration date in the second method.
The choice of either method depends on a user's preference or the limitations of authenticators.

\section{Proof of Concept and Use Cases} \label{sec:poc}
We implement the Proof of Concept (PoC) to demonstrate the feasibility that our proposal allows users to access services with multiple authenticators.
We implement the PoC using JavaScript.
In the PoC implementation, one browser window is treated as one authenticator, so that we can emulate multiple authenticators on the same device.
Note that the PoC stores seed, private keys, and the attestation key in not secure storage.
The source code is available on GitHub \footnote{\url{https://github.com/hatake5051/ovk-poc}}.

\subsection{Implementation Detail}
\subsubsection{Implementation of an Authenticator}
\begin{figure}[htb]
  \centering
  \includegraphics[width=\columnwidth]{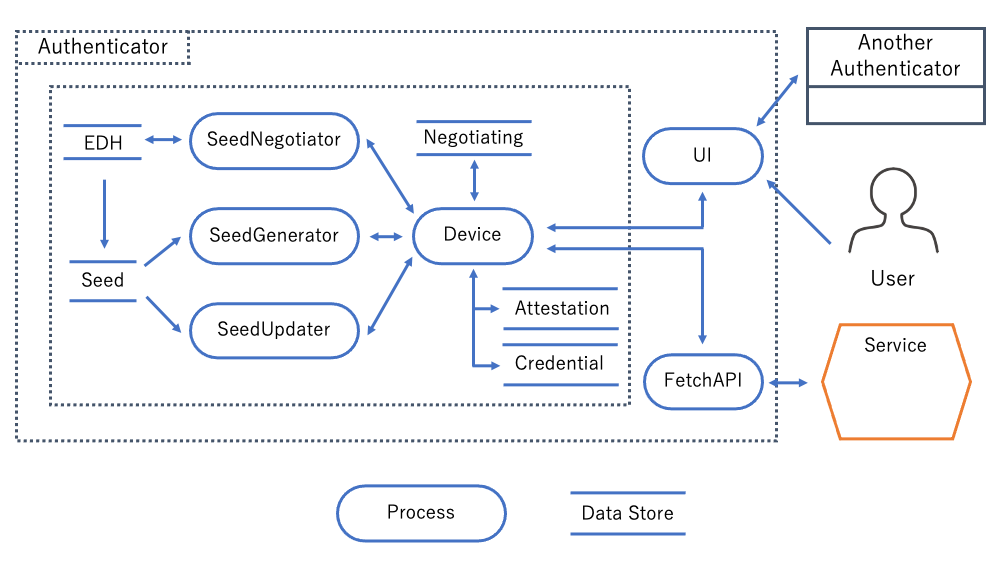}
  \caption{Authenticator Implementation}
  \label{fig:impl-authnor}
\end{figure}

Fig.\ \ref{fig:impl-authnor} illustrates the implementation of an authenticator based on data flow diagrams \cite{dfd}.
In this figure, an ellipse represents a process, an entity between an upper line and a lower line represents a data store, and an arrow represents data flow.

\texttt{SeedGenerator} implements Section \ref{sec:teian:ovk-derive}.
\texttt{Seed} stores shared seeds.
The bit length of each seed is 256.
We statically define the following parameters required in Section \ref{sec:teian:ovk-derive} and a service identifier as the origin of the service URL.

\begin{itemize}
  \item The algorithm of an OVK is elliptic curve cryptography \cite{sec1} with secp256r1 \cite{sec2}.
  An OVK is calculated using a KDF output as pseudorandomly selected an integer d of Section 3.2.1 in \cite{sec1}.
  \item The key derivation function (\texttt{KDF}) is HMAC \cite{rfc2104} using SHA-256 \cite{fips180-4}.
  \item The MAC function (\texttt{MAC}) is HMAC \cite{rfc2104} using SHA-256 \cite{fips180-4}.
\end{itemize}

\texttt{SeedNegotiator} implements Section \ref{sec:teian:seed-sharing} except for encrypting and decrypting a DH public key by a CEK and a CEK by a password, and sending and receiving ciphertexts.
\texttt{Device} implements these exceptions instead of \texttt{SeedNegotiator}.
\texttt{SeedNegotiator} stores an ephemeral private key for the DH key agreement algorithm in \texttt{EDH}.

\texttt{EDH} stores the seed calculated as a result of the key agreement in \texttt{Seed} and deletes the ephemeral private key.
We statically define the following parameters required in Section \ref{sec:teian:seed-sharing}.
We also use JSON Web Encryption Compact Serialization \cite{rfc7516} to serialize ciphertexts and algorithm identifiers.

\begin{itemize}
  \item The key agreement algorithm (\texttt{DH}) is elliptic curve diffie-hellman based on elliptic curve cryptography \cite{sec1} with secp256r1 \cite{sec2}.
  \item The authenticated encryption algorithm is AES using 128 bit key \cite{fips197} in Galois/Counter Mode (GCM) \cite{nistsp800-38d}.
  \item The password-based encryption algorithm is Password Based Encryption Scheme 2 \cite{rfc8018} using AES-KW  \cite{rfc3394} and SHA-256 \cite{fips180-4}.
\end{itemize}

\texttt{SeedUpdater} implements Section \ref{sec:teian:ovk-update}.
Only this process and \texttt{SeedGenerator} can access \texttt{Seed}.

\texttt{Device} is the process of generating key pairs and attestations and managing them based on an OVK.
\texttt{Credential} stores generated key pairs.
\texttt{Attestation} stores the attestation private key and the certificate of the corresponding attestation public key.
\texttt{Negotiating} stores data used for Section \ref{sec:teian:seed-sharing} like a password.
\texttt{Device} generates an attestation for these type of keys: public keys stored in \texttt{Credential}, OVPKs generated by \texttt{SeedGenerator}, and DH public keys calculated by \texttt{SeedNegotiator}. 

\texttt{UI} is the process of communicating ciphertexts generated by \texttt{Device} with other authenticators and interacting with a user.
We use reading and generating QR codes as the communication channel among authenticators used for Section \ref{sec:teian:seed-sharing}.
\texttt{FetchAPI} communicates with a service via a secure channel established by TLS.

\subsubsection{Implementation of a Service}
\begin{figure}[htb]
  \centering
  \includegraphics[width=0.9\linewidth]{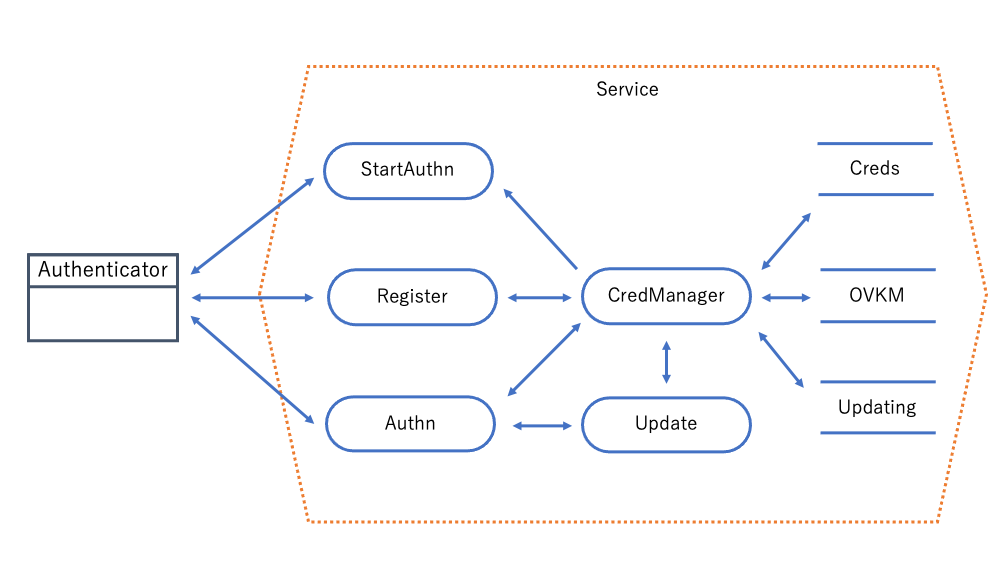}
  \caption{Service Implementation}
  \label{fig:impl-svc}
\end{figure}

Fig.\ \ref{fig:impl-svc} illustrates the implementation of a service based on data flow diagrams.

\texttt{StartAuthn} accepts authentication requests from a user.
It receives the account name of the user and responds with a challenge bound to the account and, if registered, public keys, an OVPK, and the metadata of OVPK.

\texttt{Register} accepts requests for registering a new account and a new public key bound to an account as described in Section \ref{sec:teian:ovk-derive}.
When registering a new account, a user sends a new public key, an OVPK, and the metadata of the OVPK.
The user also sends an attestation of the public key and, if requested, an attestation of the OVPK.
A service verifies attestations to determine whether it accepts the registration.

\texttt{Authn} accepts challenge responses for authentication.
We use elliptic curve digital signature algorithm \cite{sec1} with secp256r1 \cite{sec2} for authentication.
It also receives an updating message as described in Section \ref{sec:teian:ovk-update}.
\texttt{Update} processes updating OVKs.

\texttt{CredManager} manages the bindings of public keys and OVPKs to accounts.
\texttt{Creds} stores public keys bound to OVPKs.
\texttt{OVKM} stores OVPKs and the corresponding metadata bound to accounts.
\texttt{Updating} stores updating messages when processing updating OVPKs.

\subsection{Use Cases}
Using the PoC in the following scenario, we confirm that our proposal allows authenticators to share a seed, derive an OVK, register a new public key with an OVK seamlessly, and update a registered OVPK.

\begin{enumerate}
  \item Three Authenticator A, B, and C share a seed.
  \item With Authenticator A, the user registers a new public key and an OVPK with Service 1 during account registration.
  \item With Authenticator B and C, she seamlessly registers new public keys with Service 1.
  \item With Authenticator B, she registers a new public key and an OVPK with Service 2 during account registration.
  \item With Authenticator C (not A), she seamlessly registers a new public key with Service 2.
  \item Two Authenticator A and B updates a new seed (Assuming that Authenticator C is lost). 
  \item With Authenticator A, she notifies Service 1 of updating a new OVPK.
  At this time, it is still possible to sign in to Service 1 with Authenticator C.
  \item With Authenticator B, she notifies Service 1 of updating the same new OVPK as the one notified with Authenticator A.
  At this time, Service 1 updates an OVPK bound to her account, so that it is impossible to sign in to Service 1 with Authenticator C.
  \item With Authenticator B, she notifies Service 2 of updating a new OVPK.
  At this time, it is still possible to sign in to Service 2 with Authenticator C.
  After finishing the migration period, Service 2 updates an OVPK bound to her account and revokes the public key of Authenticator C.
  Then, with Authenticator A, she can seamlessly register a new public key with Service 2.
\end{enumerate}

\section{Evaluation with Threat Analysis} \label{sec:eval}
We evaluate our proposal by analyzing the PoC described in Section \ref{sec:poc} based on threat modeling \cite{threat-modeling, fido-sec-ref}.

\subsection{Security Requirement}
\subsubsection{Assets to be Protected}
We enumerate the assets to be protected in this proposal.
\begin{enumerate}
  \item Private keys stored in authenticators
  \item Public keys managed by services
  \item The attestation key stored in each authenticator
  \item The key signing attestation certificates managed by each authenticator manufacturer
  \item The trusted root certificate and policy for services to validate attestations
  \item The TLS private key managed by services
  \item The trusted root certificate and policy for authenticators to validate TLS certificates
  \item The seed stored in authenticators
  \item The OVPK and corresponding metadata stored in services
  \item The ephemeral DH private key generated by authenticators for sharing a seed
  \item The temporary password stored in authenticators for sharing a seed
\end{enumerate}

\subsubsection{Security Goals} \label{sec:eval:sec-goals}
We enumerate the goals to achive in our proposal.
We define the following goals while referring to goals in \cite{fido-sec-ref}.

\begin{description}
  \item[SG-1] Strong User Authentication: Services can authenticate users based on public key authentication.
  \item[SG-2] Unlinkability: Services cannot correlate their accounts.
  \item[SG-3] Credential Binding: Services can bind public keys to legitimate accounts.
  \item[SG-4] Attestable Properties: Services and authenticators can validate public keys by verifying attestations.
  \item[SG-5] Forgery Resistance: Be resilient from attempting to modify intercepted communications to masquerade as legitimate users.
\end{description}

\subsubsection{Security Assumptions}
We enumerate the assumptions about the environment where our proposal works.

\begin{description}
  \item[SA-1] The processes and data stores surrounded by the trusted boundary represented by the inner dotted line in Fig.\ \ref{fig:impl-authnor} are isolated from other processes on the authenticator.
  The authenticator requires local authentication before accessing these processes and data stores.
  \item[SA-2] The cryptographic algorithms used can achieve the objectives of each algorithm.
  \item[SA-3] A service can correctly validate the certificate chain of attestations.
  \item[SA-4] A service and an authenticator can establish a secure channel for service authentication, confidentiality for message, and integrity for messages (like TLS).
\end{description}

\subsection{Threat Analysis}
We enumerate the threats on data flow diagrams described in Section \ref{sec:poc}, and explain what goals listed in Section \ref{sec:eval:sec-goals} each threat violates and what measures our proposal takes.

\subsubsection{Threats on an Authenticator}
In Fig.\ \ref{fig:impl-authnor}, a dotted line represents a trusted boundary.
We focus on data flows across trusted boundaries and enumerate the threats in each of them.

First, threats arising between a user and \texttt{UI} include the following.

\begin{description}
  \item[Homograph Mis-Registration] A malicious service pretends a legitimate service to make the user believes it is legitimate.
  It prompts the user to register a new public key seamlessly, and sends metadata stolen from other services.
  The malicious service correlates OVPKs by whether the user requests a new public key registration or not.
  This threat violates SG-2 because the malicious service can correlate different OVPKs generated for different services.
  Our proposal addresses this threat because authenticators verify the MAC value of the received metadata.
  The data protected by the MAC value contains the identifier of the service that the authenticator communicates.
  Because Assumption SA-4 allows authenticators to trust the communicating service identifier, authenticators can detect spoofing of services by malicious actors.
  \item[User Verification By-pass] An attacker can operate the authenticator, or an attacker can bypass the local authentication of the authenticator to operate it.
  This threat violates SG-1 because the attacker can masquerade as the legitimate user.
  Our proposal addresses this threat by transferring it to SA-1.
  We consider the case where this SA-1 assumption is not satisfied during the OVK update in Section \ref{sec:kousatsu:update}.
\end{description}

Threats arising between a service and \texttt{FetchAPI} include the following.

\begin{description}
  \item[Service Verification Error] The authenticator cannot properly authenticate services, thus it cannot correctly identify services.
  As a result, an attacker can eavesdrop and tamper with the communication channel.
  This threat violates SG-1 and SG-5 because the attacker can hijack the authenticated session.
  This threat also violates SG-2 because the authenticator fails to address the threat Homograph Mis-Registration described above.
  Our proposal addresses this threat by transferring it to SA-4.
\end{description}

Threats arising between another authenticator and \texttt{UI} include the following.

\begin{description}
  \item[Malicious Authenticator Linking] An attacker's authenticator participates in sharing a seed.
  The attacker gets the seed value itself.
  The attacker can use this authenticator to register a new public key with any service that a legitimate user has registered.
  This threat violates SG-3.
  Our proposal addresses this threat because a user protects sharing a seed with a password.
  Users are required to use passwords that are not guessable during sharing a seed.
  \item[Weak Authenticator] A user allows a weak authenticator to participate in sharing a seed.
  The weak authenticator does not securely protect a seed, so, when an attacker compromises the weak authenticator, the seed may be leaked.
  This threat violates SG-3 because the attacker can register a new public key of the attacker's authenticator by generating the OVK with the compromised seed.
  Our proposal addresses this threat because each authenticator validates the security properties of other authenticators through attestations when sharing a seed.
\end{description}

Threats arising between \texttt{Device} and \texttt{UI} include the following.

\begin{description}
  \item[Malicious Authenticator] A user uses a malicious authenticator.
  Because a user cannot rely on the malicious authenticators, this threat violates any goals.
  It is difficult for users to determine whether it is a legitimate authenticator from a trusted manufacturer.
  Our proposal addresses this threat because services maintain a list of attestation certificates of trusted manufacturers.
\end{description}

Threats to the authenticator include the following.

\begin{description}
  \item[Side Channel Attack] Access to the data store that is not described in Fig.\ \ref{fig:impl-authnor} compromises assets to be protected.
  This threat violates SG-1 if key pairs are compromised and SG-2 and SG-3 if seeds are compromised.
  This threat also arises another threat like Malicious Authenticator Linking if a temporary password for sharing a seed is compromised.
  Our proposal addresses this threat by transferring it to SA-1.
  \item[Bad Cryptography Primitives] An authenticator uses a compromised cryptographic algorithm or a weak pseudo-random number generator in the process.
  This threat violates SG-3 if an attacker derives the OVK.
  This threat also violates SG-2 if an attacker gets the seed.
  Our proposal addresses this threat by transferring it to SA-2.
\end{description}

\subsubsection{Threat on a Service}
In Fig.\ \ref{fig:impl-svc}, we focus on data flows across trusted boundaries and enumerate the threats in each of them.

Threats arising between an authenticator and \texttt{StartAuthn} include the following.

\begin{description}
  \item[Linking OVPK] An attacker can obtain the OVPK and metadata associated with the account.
  The attacker receives OVPKs and corresponding metadata from many services and attempts to derive corresponding OVSKs.
  The attacker also attempts to correlate collected OVPKs by checking whether OVPKs are derived from the same seed.
  This threat violates SG-2 and SG-3.
  Our proposal addresses this threat by transferring it to SA-2.
\end{description}

Threats arising between an authenticator and \texttt{Register} include the following.

\begin{description}
  \item[Malicious Authenticator Registration] An attacker attempts to register a new public key of his authenticator to a legitimate user account.
  This threat violates SG-3.
  Our proposal addresses this threat because an attacker cannot get the OVSK corresponding to the OVPK bound to the account.
  An attacker cannot also get the seed corresponding to the OVPK.
  The service can verify that trusted authenticators store a seed and OVSKs by verifying OVPK attestations.
  Even if a seed is compromised, our proposal mitigates this threat because the number of authenticators that can be registered is restricted.
  After a user registers public keys of her all authenticators an attacker cannot register his public keys.
\end{description}

Threats arising between an authenticator and \texttt{Authn} include the following.

\begin{description}
  \item[Updating Malicious OVPK] An attacker attempts to update to an OVPK derived from the seed held by his authenticator.
  This threat violates SG-3 because the attacker can register his public keys.
  This threat also violates SG-1 because the attacker can revoke the user's public keys.
  Our proposal addresses this threat because an attacker cannot know the seed corresponding to the registered OVPK.
  We consider the case where this SA-1 assumption is not satisfied during updating an OVK in Section \ref{sec:kousatsu:update}.
  We disscuss whether our proposal mitigates this threat even if the seed is compromised in Section \ref{sec:kousatsu:update}.
\end{description}

\section{Discussion} \label{sec:kousatsu}
\subsection{Deriving an OVK}
We confirm that our proposal achieves the requirements of Section \ref{sec:teian:ovk-derive:youken}.
First, a user registers an OVK during new account registration where a service can trust information received from a user.
A service can verify the owner of the authenticator storing a private key corresponding to the public key to be registered without a trusted third party.
Note that attestation and the safety of communications rely on a trusted third party.
Second, malicious services cannot correlate their accounts with sharing OVPKs and corresponding metadata because the security property of a key derivation function makes it impossible to derive a seed from an OVPK and the corresponding metadata.
Besides, a malicious service cannot correlate the user's account by checking whether a user can use the OVPK and the metadata of another service to request a new public key registration.
This is because authenticators verify the MAC value of the received metadata to determine whether the received metadata is for the service.
Third, an authenticator generates an attestation of the public key to be registered.
A service can verify the attestation to evaluate the trustworthiness of the public key requested to be registered.
Finally, once authenticators of a user share a seed, they derive an OVK per service independently.
She does not have to operate multiple authenticators whenever registering a new public key.

\subsection{Sharing a Seed}
We confirm that our proposal achieves the requirements of Section \ref{sec:teian:seed-sharing:youken}.
Only authenticators having a password can participate in sharing a seed.
Only authenticators having a password can decrypt the ciphertexts generated by other authenticators having the same password.
They can also verify the integrity of received ciphertexts by a password.
A user enters a password directly into each authenticator so that the password does not flow on the communication channel where authenticators share a seed.
Because a long enough password allows an attacker to take an extremely long time to decrypt ciphertexts and a secure encryption algorithm prevents him from compromising ciphertexts, it is difficult for the attacker to compromise the password and participate in sharing a seed before the sharing is complete.
Besides, because the assumption that a DH key agreement algorithm is secure against eavesdropping prevents an attacker from deriving a seed from decrypted ciphertexts including DH public keys, an attacker cannot compromise a seed.
From the above, without assuming the security of the communication channel, we can prevent the leakage of the seed by eavesdropping and tampering.

\subsection{The Trustworthiness of a Key for Authentication}
A service considers the next two to determine whether it trusts a key for authentication.
One is the trustworthiness of the key itself.
The other is the trustworthiness of the binding of a public key to an account.
A service evaluates the former by verifying that a trusted authenticator stores a private key and that a used cryptographic algorithm has not been compromised.
To verify that, a service verifies the attestation of a public key.

A service evaluates the latter by verifying whether the private key corresponding to the public key bound to an account is stored in the authenticator owned by the user having the account.
This trustworthiness depends on how a user registers the public key.
For example, in \cite{fido-multi-registration}, a user registers a new public key via an authenticated session established by a registered public key.
In our proposal (Section \ref{sec:teian:ovk-derive} and Section \ref{sec:teian:seed-sharing}), a user registers a new public key with an OVK derived from the seed shared among authenticators.
In the former method, a public key has high assurance because a user uses registered authenticators every time she registers a new public key.
The method is not convenient because she has to have a registered authenticator for registering a new one.
The latter method (our proposal) is convenient because, once she has shared a seed, a user must have only an authenticator to append a new public key to a service.
A public key does not have high assurance if the seed can be compromised.

To make a public key higher assurance in our proposal, We propose a method for a service to verify the trustworthiness of an OVPK (Section \ref{sec:teian:ovk-trust}).
By verifying an attestation of an OVPK to be registered, a service can evaluate whether the seed deriving the OVPK is stored securely on the authenticator communicating with the service.
By verifying an attestation of a DH public key when sharing a seed, authenticators can evaluate whether the other authenticators store the seed securely.
A service can also verify whether all authenticators securely store the seed deriving the registered OVPK by verifying an attestation of the OVPK.
This is because the attestation includes the model names of all authenticators storing the seed.
A service can verify whether they store the seed securely based on the trusted policy about what authenticator model has secure storage and stores the seed in the storage.

\subsection{Updating an OVK} \label{sec:kousatsu:update}
A service calculates the trustworthiness of an updating message and selects the OVK of the most trusted message as the new OVK.
A service considers the number of registered authenticators sending the same updating message as the trustworthiness of the updating message.
We consider that the trustworthiness of all registered authenticators before the OVK migration period is equal.
This is because it is difficult for a service to determine whether an authenticator is stolen or held by a legitimate user.
Based on the assumption that it takes time for an attacker to gain control of a stolen authenticator (Assumption 2 in Section \ref{sec:teian:ovk-update:assumption}), a service selects the earlier sent message when two or more updating messages have the same and most trustworthiness. 

We discuss what attacks the proposed method prevents when an attacker can operate the seed and the private key with a stolen authenticator (Assumption 1 in Section \ref{sec:teian:ovk-update:assumption}).
In our proposal, the number of registered authenticators is limited to $N$, which is in the metadata sent when an OVK registration.
Note that an attacker can increase the number of registered authenticators by registering public keys in the way described in Section \ref{sec:teian:ovk-derive} before sending updating messages.

We disscuss cases based on the following numbers.
\begin{itemize}
  \item $N$: the number of authenticators sent when an OVK registration
  \item $N_u$: the number of registered authenticators owned by a legitimate user before an OVK migration period
  \item $N_a$: the number of registered authenticators controlled by an attacker before an OVK migration period
\end{itemize}

In the case of $N=2$, the service trusts the authenticator that sends updating messages earlier.
Therefore, in the case of $N_a =1$, if the attacker sends an updating message earlier (Assumption 2 in Section \ref{sec:teian:ovk-update:assumption} is broken), the service trusts the OVK sending from the authenticator stolen by the attacker and revokes the public key whose corresponding private key is held by the authenticator of the legitimate user.

In the case of $N \geq 3$, if $N_u \geq N / 2$ or $N_u > N_a$, then a legitimate user can update an OVK and prevent the attacker from updating an OVK because public keys whose corresponding private keys is stored in stolen authenticators are correctly revoked.

\section{Conclusion} \label{sec:conclusion}
We introduce a key pair called an Ownership Verification Key (OVK) and propose the mechanism where users and services manage public keys based on the owner of authenticators storing the corresponding private keys.
The mechanism allows users to access services with any of their authenticators without registering each of their public keys explicitly.

A user can derive the private key of an OVK (OVSK) on her authenticators from the seed sharing among the authenticators.
A service binds the public key of OVK (OVPK) to a user's account.
A service binds a public key signed by an OVSK to the user's account bound to the corresponding OVPK.
When a user changes a set of her authenticators, she updates an OVSK, and a service updates an OVPK binding to her accounts based on the most trustworthy updating message.

We implemented the Proof of Concept and confirmed that key management works as expected for typical use cases.
With threat modeling, we evaluated what measures our proposal takes against the threats.
We confirmed that our proposal achieves some security goals, such as that services cannot correlate accounts and can correctly bind public keys to accounts.
We discussed how our proposal mitigates threats for which measures are not sufficient. 

Future work includes a model where, in updating an OVK, the trustworthiness of each authenticator having the private key corresponding to a registered public key differs.
For example, when a service receives a message that some authenticator is not trustworthy from the registered email address, the service reduces the trustworthiness of the authenticator.

\bibliographystyle{ACM-Reference-Format}
\bibliography{paper}

\end{document}